\documentclass[12pt]{article}

\usepackage{amsmath,amssymb,amsthm,cite,mathtools,xcolor}
\usepackage[margin=3cm]{geometry}

\usepackage[T1]{fontenc}
\usepackage[osf]{newtxtext} 
\usepackage[nosymbolsc,cmintegrals]{newtxmath}
\usepackage[cal=euler,frak=euler]{mathalfa} 
\usepackage{bm} 

\title{On the kinematics of the last Wigner particle}

\author{J. M. Gracia-Bond\'ia$^{1,2}$ and J. C. V\'arilly$^3$
\\[18pt]
{\normalsize $^1\,$Escuela de F\'isica,
Universidad de Costa Rica, San Jos\'e 11501, Costa Rica}\\[3pt]
{\normalsize $^2\,$Departamento de F\'isica Te\'orica, Universidad de
Zaragoza, Zaragoza 50009, Spain}\\[3pt]
{\normalsize $^3\,$Escuela de Matem\'atica,
Universidad de Costa Rica, San Jos\'e 11501, Costa Rica}}

\date{August 14, 2019}

\DeclareMathOperator{\arcsinh}{arcsinh} 

\newcommand{\al}{\alpha}            
\newcommand{\bt}{\beta}             
\newcommand{\dl}{\delta}            
\newcommand{\eps}{\varepsilon}      
\newcommand{\ga}{\gamma}            
\newcommand{\ka}{\kappa}            
\newcommand{\La}{\Lambda}           
\newcommand{\om}{\omega}            
\newcommand{\sg}{\sigma}            
\renewcommand{\th}{\theta}          
\newcommand{\ze}{\zeta}             

\newcommand{\bC}{\mathbb{C}}        
\newcommand{\bJ}{\mathbb{J}}        
\newcommand{\bK}{\mathbb{K}}        
\newcommand{\bL}{\mathbb{L}}        
\newcommand{\bM}{\mathbb{M}}        
\newcommand{\bP}{\mathbb{P}}        
\newcommand{\bS}{\mathbb{S}}        
\newcommand{\bT}{\mathbb{T}}        
\newcommand{\bW}{\mathbb{W}}        

\newcommand{\sH}{\mathcal{H}}       
\newcommand{\sP}{\mathcal{P}}       

\newcommand{\pl}{\mathfrak{p}}      

\newcommand{\intl}{{\mathrm{int}}}  
\newcommand{\const}{{\mathrm{const}}} 
\newcommand{\cov}{{\mathrm{cov}}}   
\newcommand{\stan}{{\mathrm{st}}}   

\bmdefine{\aaa}{a}          
\bmdefine{\bb}{b}           
\bmdefine{\cc}{c}           
\bmdefine{\kk}{k}           
\bmdefine{\KK}{K}           
\bmdefine{\LL}{L}           
\bmdefine{\mm}{m}           
\bmdefine{\nn}{n}           
\bmdefine{\one}{1}          
\bmdefine{\PP}{P}           
\bmdefine{\pp}{p}           
\bmdefine{\RR}{R}           
\bmdefine{\ssg}{\sigma}     
\bmdefine{\SSS}{S}          
\bmdefine{\TT}{T}           
\bmdefine{\ttt}{t}          
\bmdefine{\vv}{v}           
\bmdefine{\WW}{W}           
\bmdefine{\ww}{w}           
\bmdefine{\xii}{\xi}        
\bmdefine{\xx}{x}           
\bmdefine{\YY}{Y}           
\bmdefine{\zero}{0}         

\newcommand{\del}{\partial}         
\newcommand{\up}{{\mathord{\uparrow}}} 
\newcommand{\wt}{\widetilde}        
\newcommand{\x}{\times}             
\newcommand{\7}{\dagger}            
\renewcommand{\.}{\cdot}            
\renewcommand{\:}{\colon}           

\newcommand{\half}{{\mathchoice{\thalf}{\thalf}{\shalf}{\shalf}}}
\newcommand{\ihalf}{\tfrac{i}{2}}   
\newcommand{\quarter}{\tfrac{1}{4}} 
\newcommand{\shalf}{{\scriptstyle\frac{1}{2}}} 
\newcommand{\thalf}{\tfrac{1}{2}}   

\newcommand{\ket}[1]{\,\lvert#1\rangle} 
\newcommand{\Ket}[1]{\,\bigl\lvert#1\bigr\rangle} 
\newcommand{\word}[1]{\quad\text{#1}\quad} 

\def\wick:#1:{\,\mathopen:#1\mathclose:\,} 

\def\epsi^#1_#2{\eps^{#1}{}_{\!#2}} 
\def\epsii_#1^#2{\eps_{#1}{}^{\!#2}} 

\def\lLa^#1_#2{\Lambda^{#1}{}_{\!#2}}  
\def\ltLa_#1^#2{\Lambda_{#1}{}^{\!#2}} 

\newcommand{\braket}[2]{\langle#1\mathbin|#2\rangle} 

\newcommand{\twobytwo}[4]{\begin{pmatrix} #1 & #2 \\ #3 & #4
\end{pmatrix}}

\theoremstyle{plain}
\newtheorem{thm}{Theorem}           

\theoremstyle{definition}
\newtheorem{remk}[thm]{Remark}      

\makeatletter
\renewcommand{\section}{\@startsection{section}{1}{\z@}%
                       {-3.5ex \@plus -1ex \@minus -.2ex}%
                       {2.3ex \@plus.2ex}%
                       {\normalfont\large\bfseries}}
\renewcommand{\subsection}{\@startsection{subsection}{2}{\z@}%
                       {-3.25ex \@plus -1ex \@minus -.2ex}%
                       {1.5ex \@plus .2ex}%
                       {\normalfont\normalsize\bfseries}}
\makeatother

\numberwithin{equation}{section}

\begin{document}

\maketitle

\begin{abstract}
Wigner's particle classification provides for ``continuous spin''
representations of the Poincar\'e group, corresponding to a class of
(as yet unobserved) massless particles. Rather than building their
induced realizations by use of ``Wigner rotations'' in the textbooks'
way, here we exhibit a scalar-like first-quantized form of those
(bosonic) Wigner particles directly, by combining wave equations
proposed by Wigner long ago with a recent prequantized treatment
employing Poisson structures.
\end{abstract}

\section{Introduction}
\label{sec:introibo}

By the last Wigner particle (WP) here is meant the last case in
Wigner's classification of unirreps of the Poincar\'e
group~\cite{Wigner39}: massless particles whose second Casimir has a
nonzero value. More often, they are referred to as continuous spin
particles (CSP) -- somewhat of a misnomer. Though routinely dismissed
as ``unobserved'' in standard textbook treatments, the possible
existence and properties of such particles are of continued
interest~\cite{Schroer17}; after pioneering work by Schuster and Toro
\cite{SToro13a,SToro13b,SToro15}, several recent studies
\cite{BekaertNS16,BekaertS17,Najafizadeh18,KhabarovZ18,BuchbinderKT18}
have appeared. Closer to the spirit of this paper is the construction
by Rehren~\cite{Rehren17} stemming from his own work with Mund and
Schroer -- see~\cite{MundRS17} and references therein -- of a
string-local quantum field for such a particle, as a
``Pauli--Luba\'nski limit'' of massive, string-local fields. At an
opposite end, mathematically speaking, our own construction
\cite{Euterpe} of a ``classical elementary system'' for the WP
foreshadows its quantum kinematics.

Our goal here is to review the first-quantized description of the
(bosonic) WP: this is the relevant approach for certain applications
that do not require a full-blown quantum field formalism. In
principle, such a description is already available, by means of
little-group techniques~\cite{McKerrell65,LomontM67}. However, one can
attain a simpler-looking scalar-like version by starting directly
from the wave equations. Among our purposes here is to delineate this
version, less cumbersome than the standard approach.

The plan of the article is as follows. In Section~2 we recall the
theory of the second Casimir associated to the Poincar\'e group,
borrowing a method and notation going back to work by
Schwinger~\cite{Schwinger70}. We also bring in a quite instrumental
result on the Wigner rotation for massless particles~\cite{Ganymede}.
Section~3 is the core of the paper. There we introduce an appropriate
set of states for the WP, and we show the invariant nature of their
associated wavefunctions, their equations of motion, and the existence
of an invariant scalar product. In Section~4 we exhibit the causal
propagator for the boson WP. Section~5 deals briefly with the relation
between the invariant and the conventional formalisms.

In the appendices we state and develop our Poincar\'e-group
conventions, and then expound a relevant aspect of little-group theory
that we have not found in the standard presentations.

\section{The Schwinger decomposition of the Pauli--Luba\'nski
operator}
\label{sec:PLanning-stage}

Before coming to the (one-particle) Hilbert space for the WP, let us
recall the standard basis of the Lie algebra $\pl$ of the Poincar\'e
group $\sP_+^\up$ whose $10$~generators 
$\{P^0,P^a,L^a,K^a : a = 1,2,3\}$ correspond respectively to time
translation, space translations, rotations and boosts -- consult
Appendix~\ref{app:back-to-basics} for our notation and conventions.
The commutation relations for the Lorentz subgroup are as follows:
$$
[L^a,L^b] = \epsi^{ab}_c\, L^c,  \qquad 
[L^a,K^b] = \epsi^{ab}_c\, K^c,  \qquad
[K^a,K^b] = -\epsi^{ab}_c\, L^c.
$$

The pseudovector operator (in the enveloping algebra of~$\pl$) 
\begin{align}
W^\rho &:= {J^*}^{\rho\mu} P_\mu = P_\mu {J^*}^{\rho\mu}
= (\PP \. \LL,\, P^0 \LL + \KK \x \PP) \equiv (W^0, \WW)
\notag \\
&= \bigl( P^1J^{23} + P^2J^{31} + P^3J^{12}, 
P^0J^{23} + P^2J^{30} + P^3J^{02}, 
\notag \\
&\qquad P^0J^{31} + P^1J^{03} + P^3J^{10},
P^0J^{12} + P^2J^{10} + P^1J^{02} \bigr)
\label{eq:PLot-develops} 
\end{align}
is referred to as the \textit{Pauli--Luba\'nski vector}. It clearly 
satisfies
$$
(WP) = 0  \word{and}  [P^\nu, W^\mu] = 0,
$$
and is a vector under the action of the Lorentz group generators:
$$
[J^{\mu\nu}, W^\tau] = g^{\tau\nu} W^\mu - g^{\tau\mu} W^\nu.
$$

As a corollary, one obtains the identities:
\begin{equation}
[W^\mu, W^\nu] = \epsi^{\mu\nu}_{\tau\rho} W^\tau P^\rho;  \word{and}
[J^{\mu\nu}, (WW)] = 0;
\label{eq:PLot-thickens} 
\end{equation}
the second one indicating that $(WW)$ is a Casimir operator 
for~$\sP_+^\up$. One finds also that
\begin{align}
(WW) = \quarter \eps^{\rho\mu\nu\tau} P_\mu J_{\nu\tau}\,
\eps_{\rho\ka\sg\eta} P^\ka J^{\sg\eta} 
= -\half J_{\nu\tau} J^{\nu\tau} P^2 
+ J_{\ka\sg} J^{\mu\sg} P^\ka P_\mu \,.
\label{eq:whats-what} 
\end{align}

We assume in what follows that $P^0 > 0$. By invoking expression
\eqref{eq:PLot-develops} in the rest frame, it becomes clear that the
Casimir $(WW)$ for a massive particle equals $-m^2 \SSS \. \SSS$,
where $\SSS$ is the spin generator. This tells us that $(WW)$ captures
\textit{internal} angular momentum. In general $W$ is spacelike,%
\footnote{Since $(WP) = 0$ and $(PP) \geq 0$ together imply that
$(WW) \leq 0$.}
except that in the massless case it can be parallel to~$P$: this leads
to the known fixed-helicity particles, like the photon and graviton,
for which relations~\eqref{eq:PLot-thickens} are trivial.

Here we put that case aside: the Wigner particle \textit{by
definition} obeys
$$
(WW) = -\ka^2 < 0.
$$
We have seen in \eqref{eq:PLot-develops} that the temporal component
of~$W$ is directly related to helicity, which deserves a symbol:
\begin{align}
H := (\PP \. \LL)/P^0. \word{Therefore} W^0 = H P^0.
\label{eq:enter-helicity} 
\end{align}
The relation $(WP) = 0$ implies that the relevant part of $\WW$ is
that which is transverse to~$\PP$:
\begin{equation}
\TT := \WW - W^0\PP/P^0 = \WW - (\WW\.\PP)\PP/(P^0)^2,
\word{so that}  \WW = H\PP + \TT.
\label{eq:tea-break} 
\end{equation}
Notice that $\TT^2 = \ka^2$. We call $\WW = H\PP + \TT$ the Schwinger
decomposition of (the spatial part of) the PL vector; the notation
$\TT$ for the part of~$\WW$ transverse to~$\PP$ follows
Ref.~\cite{Schwinger70}. Not only do the components of $\TT$ commute
with the momentum; they commute with each other. This is worth a
proof:
\begin{align*}
[T^a, T^b] &= [W^a, W^b] - [W^0, W^b] P^a/P^0 - [W^a, W^0] P^b/P^0
\\
&= \epsi^{ab}_c\, T^c P^0 - \epsi^b_{de} T^d P^e P^a/P^0 
+ \epsi^a_{rs} T^r P^s P^b/P^0 
\\
&= \epsi^{ab}_c\, T^c P^0 - (\TT \x \PP)^b P^a/P^0
+ (\TT \x \PP)^a P^b/P^0
\\
&= \epsi^{ab}_c \bigl( T^c P^0 + ((\TT \x \PP) \x \PP)^c/P^0 \bigr)
= \epsi^{ab}_c \bigl( T^c P^0 - T^c(P^0)^2/P^0 \bigr) = 0.
\end{align*}
Schwinger writes for this: $\TT \x \TT = \zero$. Note also that
$$
[H, K^a] = T^a/P^0; \qquad [K^a, T^b] = T^a P^b/P^0.
$$

Let us introduce another spatial $3$-vector, also transverse to~$\PP$:
$$
\YY := (\PP/P^0) \x \TT.
$$
There is a $4$-vector naturally associated with~$\YY$ like $W$ with 
$\TT$. But we do not go into that. Note the commutator relation
$$
[H, T^a] = [W^0, T^a]/P^0 = \epsi^a_{bc}\, T^b P^c/P^0
= \TT \x \PP^a/P^0 = - Y^a,
$$
which is at once accompanied by
\begin{align*}
[H, Y^a] &= \epsi^a_{bc} [H, P^b T^c/P^0] = -\epsi^a_{bc} P^b Y^c/P^0
\\
&= -(\PP \x \YY)^a/P^0 = (\PP \x (\TT \x \PP))^a (P^0)^{-2} = T^a.
\end{align*}

At this point, following Schwinger anew, and also inspired by
\cite{BalachandranMSSZ92}, we may introduce a position vector
commuting with~$H$:
$$
\RR = -\half [\KK, (P^0)^{-1}]_+ - (\TT \x \PP)(P^0)^{-3}.
$$
Notice that
$$
[W^0, \RR] = - \frac{\WW_\parallel}{P^0}\,,  \word{so}
[H, \RR] = [W^0/P^0, \RR]
= - \frac{\WW_\parallel}{(P^0)^2} + \frac{W^0\PP/P^0}{(P^0)^2} = 0.
$$
We remark that $[P^j, R^k] = - \dl^{jk}$. Also,
$[R^j, P^0] = P^j/P^0$ and $[R^j, (P^0)^{-1}] = -P^j/(P^0)^3$.

We list here some commutators involving $\RR$:
\begin{align}
& [R^j, P^k] = \dl^{jk},  \quad
[R^j, P^0] = P^j(P^0)^{-1},  \hspace{2.55em}
[W^0, R^j] = (T^j - W^j) (P^0)^{-1},
\notag \\
& [R^j, H] = 0, \,\qquad [R^j, T^k] = - T^j P^k (P^0)^{-2}, \quad
[R^j, R^k] = - \eps^{jk}_l H P^l (P^0)^{-3},
\notag \\
& [R^j, (\TT\x\PP)^k(P^0)^{-2}] + [(\TT\x\PP)^j(P^0)^{-2}, R^k] = 0.
\label{eq:non-loco} 
\end{align}
The sixth relation in~\eqref{eq:non-loco} shows the WP to be
intrinsically non-localizable. The proofs of the above are routine;
and anyway, the Poisson brackets and general results of the thorough
study of the kinematics of the WP in Kirillov's prequantized
formalism~\cite{Euterpe} can be largely transposed here. In
particular: the commuting orthogonal trihedron $(\PP, \TT, \YY)$
\textit{rotates gyroscopically} under boosts, this being \textit{ipso
facto} true for all (restricted) Lorentz transformations. While the
length of $\PP$ can vary, the lengths of $\TT$ and~$\YY$ are fixed
at~$\ka$. The next subsection helps to understand why.

\subsection{The Wigner rotation, tamed}
\label{sec:Wignerian-rot}

In the massive case there is a canonical definition for a Lorentz
transformation taking the reference momentum $(m,\zero)$ to $p$, as a
boost $L_{\ze\nn}$ with direction $\nn$ (a unit vector) and boost
parameter~$\ze$. The corresponding Wigner rotation acting%
\footnote{In the ``active transformation'' view
\cite[Sect.~3.3]{SexlU01}.}
on a $3$-vector $\vv$ is found in~\cite{Euterpe,Ganymede}:
\begin{align*}
R(L_{\ze\nn},p) \vv = R_{\mm,\dl} \vv 
&= \vv\,\cos\dl + \mm\x\vv \,\sin\dl + (\mm\.\vv) \mm (1 - \cos\dl),
\\
\shortintertext{where:}
\mm = \frac{\pp \x \nn}{|\pp \x \nn|}; \quad 
\cos\dl &= 1 - \frac{|\pp\x\nn|^2(\cosh\ze - 1)}{(m + p^0)(m + p'^0)},
\\[\jot]
\sin\dl &= \frac{(m + p^0)\sinh\ze + \nn\.\pp(\cosh\ze - 1)}
{(m + p^0)(m + p'^0)}\, |\pp \x \nn|,
\end{align*}
with the action $p \mapsto p'$ on $4$-momenta given by:
\begin{align*}
p'^0 &= p^0 \cosh\ze + \nn\.\pp \sinh\ze,
\\
\pp' &= \pp + p^0\,\nn\,\sinh\ze + (\nn\.\pp) \nn (\cosh\ze - 1).
\end{align*}

As remarked in~\cite{Ganymede}, the massless limit of $\sin\dl$ is
perfectly smooth:
\begin{equation}
\sin\dl = \biggl( \frac{\sinh\ze}{p'^0} 
+ \frac{\nn\.\pp(\cosh\ze - 1)}{p^0 p'^0} \biggr) |\pp\x\nn|,
\label{eq:Wigner-angle} 
\end{equation}
whereas
$$ 
\pp \x \pp' 
= \bigl[ p^0 \sinh\ze + \nn\.\pp (\cosh\ze - 1) \bigr] \pp \x \nn;
$$
therefore the component of $\pp'$ not along $\pp$ stays in the plane
perpendicular to $\pp \x \nn$. The sine of the angle of rotation is
given by
\begin{equation}
\frac{|\pp \x \pp'|}{|\pp||\pp'|} 
= \frac{p^0 \sinh\ze + \nn\.\pp(\cosh\ze - 1)}{|\pp||\pp'|}
\,|\pp \x \nn|.
\label{eq:rotation-angle} 
\end{equation}
In the massive case (where $p^0 p'^0 > |\pp||\pp'|$), this angle is
generally greater than the Wigner rotation angle~$\dl$. The key point
is that this formula makes perfect sense for $m = 0$, even though some
of the factors in its definition do not. Namely, keeping in mind that
in the massless case $p^0 = |\pp|$ and $p'^0 = |\pp'|$, the
formula~\eqref{eq:rotation-angle} exactly matches
formula~\eqref{eq:Wigner-angle}. Which means that momentum and
``spin'' turn in solidarity. Wigner graphically describes why in the
massless case they must do so: ``for a particle with zero rest-mass
[\dots] if we connect any internal motion with the spin, this is
perpendicular to the velocity''~\cite{Wigner57}.

\section{The invariant formalism for the WP}
\label{sec:unknown-master}

To construct a Hilbert space $\sH$ carrying a unitary irreducible
representation (or ``unirrep'') $U$ of the Poincar\'e group
$\sP_+^\up$ corresponding to a Wigner particle with Casimir~$\ka^2$,
we proceed by taking a basic set of kets, labelled as
$\Ket{|\pp|,\pp/|\pp|,\ttt}$; where%
\footnote{We use open-faced type for the operators on Hilbert space
corresponding to geometrical generators.}
$$
\bP^\mu \ket{\pp,\ttt} = \bP^\mu \Ket{|\pp|, \pp/|\pp|, \ttt}
= p^\mu \Ket{|\pp|,\pp/|\pp|,\ttt}; \qquad
\bT \Ket{|\pp|, \pp/|\pp|, \ttt} = \ttt \Ket{|\pp|, \pp/|\pp|, \ttt}.
$$
Here $\bP^\mu$ is the selfadjoint operator corresponding to the
generator $P^\mu$; $\bT$ is the $3$-component selfadjoint operator
corresponding to Schwinger's geometric generator~$\TT$; and $\ttt$ is
the \mbox{$3$-vector} of its eigenvalues. These polarization states
lie on a circle of radius~$\ka$ in the plane perpendicular to~$\pp$.
Thus, with some abuse of notation, we can rewrite $\ket{\pp,\th}$ or
$\ket{\ka;\pp,\th}$ for those kets, with $\th$ denoting their angular
degree of freedom. Note that different positive values of~$\ka$
correspond to inequivalent representations of~$\sP_+^\up$\,.

The gyroscopic property is the key to the strange simplicity of the WP
structure, as it indicates that the corresponding wave-functions for
the WP may transform similarly to spin-zero particles. Indeed, for
\textit{any} Lorentz transformation $\La$ the gyroscopic property
implies that the rotation $R_\La \: \pp/|\pp| \mapsto \pp'/|\pp'|$
applies equally to~$\ttt$, i.e., $\ttt \mapsto \ttt' = R_\La \ttt$.
This is clear if $\La$ is a rotation, and has been shown
in~\cite{Euterpe} when $\La$ is a boost; and so it is true of
any~$\La$.

\begin{remk} 
The little-group techniques demand the choice of a Lorentz
transformation at each point of (the mantle of) the lightcone. Now, it
is not possible, for rather obvious topological
reasons~\cite{BoyaCS74}, to construct a global continuous section of
the $SL(2,\bC)$-principal bundle. Since one works mostly in the
category of Hilbert spaces, and there \emph{exist} Borel sections,
this is usually deemed not too serious a problem. However, it does
produce some pathologies, which, according to the analysis
in~\cite{FlatoSF83}, for ordinary massless particles of nonzero
helicity at least, partially invalidate the concept of sharp momentum
states that people have been using all along. It would be good to know
whether related troubles manifest themselves for WPs in the invariant
formulation. On the other hand, the very fact that the description of
one of their states requires three angles instead of two makes for
more singular eigenstates than for scalar particles.
\end{remk}

It pertains to declare the normalization of our kets. We decide for the
Lorentz-invariant expression:
\begin{align*}
\braket{\pp,\ttt}{\pp',\ttt'}
&= |\pp| \,\dl(\pp - \pp') \,\dl(\ttt - \ttt'), \word{or}
\\
\braket{\pp,\th}{\pp',\th'}
&= |\pp| \,\dl(\pp - \pp') \,\dl(\th - \th').
\end{align*}
Let $\Phi(\pp,\th) := \braket{\pp,\th}{\Phi}$. An inner product for
these wavefunctions is thus given by
\begin{equation}
\braket{\Phi}{\Phi} 
\propto \int\frac{d^3p}{|\pp|}\,d\th\,\bigl|\Phi(\pp,\th)\bigr|^2.
\label{eq:BW-scalar-product} 
\end{equation}
The definition does not depend on the Lorentz frame
\cite{BargmannW48}. We give an explicitly invariant form of
$\braket{\Phi}{\Phi}$ in momentum space at the end of this section;
and also a formula in configuration space. In order to see them, and
to better grasp the kinematics of the WP, we introduce, following
Wigner, its manifestly invariant formalism.

\subsection{Equations of motion}
\label{sec:eppur-si-muove}

As advertised, the gyroscopic property implies that equations of
motion for the WP may be of scalar-like form. In fact, Wigner returned
many times \cite{BargmannW48,Wigner48,Wigner63} to the question of
equations of motion for a~WP. In those papers Wigner considers scalar
wave functions depending on configuration or momentum-space variables
and an extra spacelike $4$-vector variable,%
\footnote{Here called $w$, since it will be seen to be an avatar of
the PL vector.}
transforming covariantly under the Lorentz group, and satisfying the
equations:
\begin{subequations}
\label{eq:WP-wave-eqns} 
\begin{alignat}{2}
\square_x \Phi(x,w) &= 0;  \word{or}
& p^2\,\Phi(p,w) &= 0,
\label{eq:WP-wave-one} 
\\
(w^2 + \ka^2)\,\Phi(x,w) &= 0;  \word{or}
& (w^2 + \ka^2)\,\Phi(p,w) &= 0,
\label{eq:WP-wave-two} 
\\
(w\,\del_x)\,\Phi(x,w) &= 0;  \word{or}
& (pw)\,\Phi(p,w) &= 0,
\label{eq:WP-wave-three} 
\\
\bigl( (\del_x\del_w) + 1 \bigr) \Phi(x,w) &= 0;  \word{or}
& \bigl( (p\,\del_w) + i \bigr) \Phi(p,w) &= 0.
\label{eq:WP-wave-four} 
\end{alignat}
\end{subequations}

The first three equations have a ready interpretation, corresponding
respectively to the Klein--Gordon equation for a massless particle,
the value of the second Casimir associated to a given~WP, and mutual
perpendicularity of the momentum and PL vectors.

For the fourth equation, just note that identifying the equations of
motion with the action of the Casimir operators is a matter of
principle. So let us formally take $P$ and $W$ as independent
variables at the same title, in a representation in which $P$ is
diagonal, and compute from equation~\eqref{eq:whats-what} with 
$P^2 = 0$ the second Casimir:
\begin{align}
C_2 \equiv (WW) 
&= (w_\nu\,\del^w_\rho - w_\rho\,\del^w_\nu)
(w^\nu\,\del_w^\sg - w^\sg\,\del_w^\nu) \del^x_\sg \del_x^\rho
\notag \\
&= -\ka^2(\del_x\del_w)^2 + (w\,\del_x)(\del_x\del_w)
- (w\,\del_x)(w\,\del_w)(\del_x\del_w) - (w\,\del_w) \square_x
\notag \\
&\qquad - (w\,\del_x)(w\,\del_w)(\del_x\del_w)
- 4(w\,\del_x)(\del_x\del_w) 
+ (w\,\del_x)^2\,\square_w + (w\,\del_x)(\del_w \del_x)
\notag \\
&= -\ka^2 (\del_x\del_w)^2 + (w\,\del_x)^2\,\square_w
- 2(w\,\del_x)(\del_x\del_w)(w\,\del_w) - (w\,\del_w) \square_x
\notag \\
&= \ka^2(p\,\del_w)^2 - (pw)^2\,\square_w
+ 2(pw)(p\,\del_w)(w\,\del_w) = - \ka^2.
\label{eq:papa-de-los-tomates} 
\end{align}
Now, since here $(pw) = 0$, we are left with
$(\del_x\del_w) = \mp 1$, which arguably completes the Wigner
equations \eqref{eq:WP-wave-eqns} above.%
\footnote{For definiteness, we opted for the upper sign in
\eqref{eq:WP-wave-four}; taking the lower one amounts to changing the
sign of~$\ka$ only.}

The weak point of the argument appears to be that the components of
$W$ do not commute in general. But the equations defend themselves
very well: the last one is immediately integrated,
\begin{align}
\Phi(\pp,w - \ga p) = e^{\pm i\ga}\,\Phi(\pp,w),
\label{eq:calipers} 
\end{align}
and may be interpreted as an infinitesimal gauge transformation,
which, in view of the Schwinger decomposition
\eqref{eq:enter-helicity} and \eqref{eq:tea-break}, identifies $\ga$
as the placeholder for helicity. One recognizes that the argument~$w$
in~\eqref{eq:WP-wave-eqns} stands for both ``spin'' and ``gauge''
degrees of freedom.

The Wigner system of equations is consistent; indeed, compatibility
between the third and fourth equations is guaranteed precisely by the
wave equation~\eqref{eq:WP-wave-one}, and compatibility between the
second and fourth by the third equation~\eqref{eq:WP-wave-three}. That
is to say: the differential operators in the left column
of~\eqref{eq:WP-wave-eqns} form a closed system, since $\square_x$
commutes with the other three, which have the nontrivial commutation
relations:
$$
[(\del_x\del_w) + 1, w^2 + \ka^2] = 2(w\,\del_x), \quad
[(w\,\del_x), (\del_x\del_w) + 1] = \square_x \,.
$$
This would not hold were $m > 0$, requiring $\square_x + m^2$
in~\eqref{eq:WP-wave-one}. Moreover, were $\ka = 0$, then
\eqref{eq:WP-wave-four} would not follow from
\eqref{eq:papa-de-los-tomates}. What is more: in the light of the
display above, the two key equations are \eqref{eq:WP-wave-four} and
\eqref{eq:WP-wave-two}, since we may regard the other two -- whose
physical meaning is obvious -- as their compatibility conditions. In
summary: the system \eqref{eq:WP-wave-eqns} is associated specifically
to the~WP.

Let us consider the transformation $\del_w \mapsto iv$, 
$w \mapsto -i\del_v$ in the Wigner system of equations
\cite{BekaertM06,KHRprivate}. There ensues the relation
\begin{align*}
(WW) 
&= 2(pw)(p\,\del_w)(w\,\del_w) - w^2(p\,\del_w)^2 - (pw)^2\,\square_w
\\
&= 2(pv)(p\,\del_v)(v\,\del_v) - v^2(p\,\del_v)^2 - (pv)^2\,\square_v.
\end{align*}
Therefore $(WW)$ is Fourier-invariant in this sense. 

\goodbreak 

In terms of this Fourier-conjugate to $w$, we now obtain the ``smooth
solutions'' by Schuster and Toro~\cite{SToro13a}:
$$
(p\,\del_v) \wt\Phi(\pp,v) = 0.
$$
Also, the equations in~\cite{Rehren17} coincide essentially with those
of~\cite{SToro13a}.%
\footnote{~``\dots\ alle diese Gleichungssysteme, sofern sie
widerspruchsfrei sind, \"aquivalent sind''~\cite{Wigner48}.}
The associated action functionals \cite{SToro13c,SToro15,Rivelles17}
look quite complicated.

\subsection{Invariant wavefunctions}
\label{sec:hic-jacet-lepus}

The Wigner equation~\eqref{eq:WP-wave-one} tells us that we are
on-shell in momentum. We express this by
\begin{align*}
\Phi(x,w) 
&\propto \int d^4p\, \th(p^0)\,\dl(p^2) e^{-i(px)} \Phi(\pp,w)
\propto \int\frac{d^3\pp}{|\pp|}\, e^{-i(px)} \Phi(\pp,w),
\\
\shortintertext{and equivalently}
\Phi(\pp,w) 
&\propto \int d^4x\, e^{i(px)}\, \Phi(x,w)\, \Bigr|_{p^0=|\pp|},
\end{align*}
with our choice of sign for~$p^0$. Now we may relate the above
$\braket{\pp,\ttt}{\Phi}$ with $\Phi(x,w)$. Consider again equation
\eqref{eq:WP-wave-four}, or formula \eqref{eq:calipers}, and let the
gauge $\ga := w^0/p^0 = w^0/|\pp|$. It follows that
\begin{align*}
\Phi(\pp,w) &\equiv \Phi(\pp,w^0,\ww_\parallel,\ttt)
= \exp(-iw^0/|\pp|)\, \Phi(\pp,0,\ttt)
\\
&=: \exp(-iw^0/|\pp|)\, \braket{\pp,\ttt}{\Phi}
= \exp(-i\,\pp\.\ww/|\pp|^2)\, \braket{\pp,\ttt}{\Phi}
\\
&=: \exp(-i(\pp\.\ww)/|\pp|^2)\, \braket{\pp,\th}{\Phi}.
\end{align*}
For any $(\pp,\th)$ there holds 
$|\Phi(\pp,\ga,\th)| = |\Phi(\pp,0,\th)|$. Notice that for the
definition \eqref{eq:BW-scalar-product} of the scalar product one
should not integrate on the real gauge variable~$\ga$, which would
yield a divergent expression.

The corresponding representation $U$ of~$\sP_+^\up$ satisfies
$$
U(a,\La)\,\Phi(x,w) = \Phi\bigl( \La^{-1}(x - a), \La^{-1}w \bigr)
$$
on the space of solutions of the equations \eqref{eq:WP-wave-eqns}. We
have found the simple theory of an invariant object for the WP -- with
the help of the Wigner equations themselves.

The internal parts of Lorentz group generators in this formalism
commute with the orbital parts. They are of the form
\cite{BargmannW48}:
\begin{align*}
\bK_{\intl,\cov}^c
&= i\bigl( w^0 \del_{w^c} + w^c \del_{w^0} \bigr) =: \bK_w^c;
\\
\bL_{\intl,\cov}^c
&= -i \epsi^c_{ab} w^a \del_{w^b} =: \bL_w^c \equiv \bS^c.
\end{align*}
Note the commutation relations $\bS \x \bS = i\bS$, in Schwinger's
notation; and that the total angular momentum generators can be
written as $\bL = -i\pp \x \del_\pp + \bS$, just like for massive
particles.

\begin{remk} 
Given $p$ such that $p^2 = 0$ and $p^0 > 0$, its three-dimensional
little group $G_p$ of rotations around $\pp/|\pp|$ and \textit{null
rotations} preserving $\pp$ is well known. Any proper, orthochronous
Lorentz transformation of the sphere must have (properly counted) two
fixed points~\cite{PenroseR84}. One possibility is that both null
directions \textit{coincide}; these are precisely the parabolic
Lorentz transformations, called in context ``null rotations''; they
are discussed further in App.~\ref{app:nullius-in-verba}.%
\footnote{The most general transformation fixing a null direction
decomposes into a null rotation (belonging to a two-parameter set), a
rotation and a boost. The four of them together constitute a
\textit{Borel subgroup} of the Lorentz group; the last two have as
invariant directions those of $\kk$ and the antipodal $-\kk$; the
boost does not leave $k$~itself invariant.}

Given a pair $(p,w)$ satisfying $p^0 > 0$, 
$p^2 = (pw) = w^2 + \ka^2 = 0$ and another pair $(p',w')$ of the same
kind, there is a \textit{unique} restricted Lorentz transformation
$\La$ such $\La p = p'$ and $\La w = w'$.
\end{remk}

\begin{remk} 
The scalar product \eqref{eq:BW-scalar-product} is Lorentz-invariant,
though not obviously so. A manifestly invariant form of the scalar
product appears in Wigner~\cite{Wigner48}: given two solutions
$\Phi(p,w)$, $\Psi(p,w)$ of~\eqref{eq:WP-wave-eqns}, define
$\braket{\Psi}{\Phi}$ by:
\begin{align}
2 \int d^4p \,d^4w \,\Psi^*(p,w) \,\Phi(p,w)
\,\dl(p^2) \,\dl(w^2 + \ka^2) \,\dl\bigl( (pw) \bigr) \,(pu)
\,\dl\bigl( (uw) - a \bigr),
\label{eq:nunc-pro-tunc} 
\end{align}
where $u$ is \textit{any} timelike $4$-vector such that $u^2 = 1$ and
$a$ an arbitrary parameter. For the convenience of the reader we
follow Wigner in verifying that the integral is independent of
such~$u$ and~$a$. Differentiating first with respect to~$a$,
\begin{align*}
\frac{d}{da} \braket{\Psi}{\Phi}
&= -2 \int d^4p \,d^4w \,\Psi^*(p,w)
\,\Phi(p,w) \,\dl(p^2) \,\dl(w^2 + \ka^2) \,\dl\bigl( (pw) \bigr)
\,(p\del_w) \,\dl\bigl( (uw) - a \bigr)
\\
&= 2 \int d^4p \,d^4w \,\Psi^*(p,w) \,\Phi(p,w) \,p^2\dl(p^2)
\,\dl(w^2 + \ka^2) \,\dl'\bigl( (pw) \bigr) \,\dl\bigl( (uw) - a
\bigr) = 0.
\end{align*}
Thus one can as well drop $a$ in the
expression~\eqref{eq:nunc-pro-tunc}. Next, by application of the
differential operators 
$u_\al\,\del/\del u_\bt \mp u_\bt\,\del/\del u_\al$, one easily checks
that the same expression is independent of the direction of~$u$. So we
can as well choose $u = (1,\zero)$, leading to
\begin{align*}
\braket{\Psi}{\Phi} 
&= 2 \int d^4p \,d^3\ww \,\Psi^*(p,w) \,\Phi(p,w) \,
p^0 \,\dl(p^2) \,\dl\bigl( |\ww|^2 - \ka^2 \bigr) \,\dl(\pp\.\ww)
\\
&= \int d^3\pp \,d^3\ww \,\Psi^*(p,w) \,\Phi(p,w)
\,\dl\bigl( |\ww|^2 - \ka^2 \bigr) \,\dl(\pp\.\ww),
\end{align*}
which, with $p^0 = |\pp|$ and $w^0 = 0$ in the arguments of the
wavefunctions understood, coincides with~\eqref{eq:BW-scalar-product}.

Wigner~\cite{Wigner48} discusses as well in great detail the passage
to $x$-space, yielding several equivalent forms, among which an 
attractive one is given by:
$$
\braket{\Psi}{\Phi} = \int d^3\xx \,d^3\ww \,\del_t\Psi^*(x,w)
\,\del_t\Phi(x,w) \,\dl\bigl( |\ww|^2 - \ka^2 \bigr) \,\dl(\xx\.\ww).
$$
\end{remk}

\bigskip

\section{The propagator}
\label{sec:propagapropaga}

In our notation, and with slightly different conventions, the
following formula is found in \cite[Eq.~(3.15)]{Hirata77}:
\begin{align*}
& \wt D(x,x';w^0,\ww,w'^0,\ww') = -\wt D(x',x;w^0,\ww,w'^0,\ww')
= \dl(w^2 + \ka^2)
\\
&\enspace \x \frac{1}{(2\pi)^3} \int d^3\pp \,
\frac{\sin|\pp|(t - t')}{|\pp|}\, e^{i\pp\.(\xx - \xx')} \,\dl(pw) 
\,\dl^3\bigl( |\pp|(\ww - \ww') - (w^0 - w'^0)\pp \bigr)\,
e^{i(w^0 - w'^0)/|\pp|}.
\end{align*}
The above $\wt D$ is a Lorentz invariant distribution, which 
satisfies the Wigner equations. 

Consider the skewsymmetric form $s$ given by
$$
s(\Psi,\Phi) := \int d^3x'\,\bigl[ \Psi(x')\,\del_{t'}\Phi(x')
- \Phi(x')\,\del_{t'}\Psi(x') \bigr]_{t'=\const}\,.
$$
If $D$ denotes the ordinary Jordan--Pauli propagator for massless
fields, the solution of the wave equation with Cauchy data
$\Phi(t',\xx')|_{t'=\const}$ is given by 
$s\bigl( D(x,-), \Phi(-) \bigr)$.

Now it should be clear that
\begin{align*}
\int d^4w'\, s\bigl( \wt D(x,-;w,w'), \Phi(-;w') \bigr)
&= \frac{\dl(w^2 + \ka^2)}{(2\pi)^3} 
\int d^3\ww'\, \dl^3(\ww - \ww')\, \Phi(x;\ww',w_0) \,\dl(pw)
\\
&= \Phi(x;w),
\end{align*}
if $\Phi$ already satisfies the Wigner equations; and this expression
\textit{becomes} a solution in the general case -- since $\wt D$
itself satisfies them. Therefore this $\wt D$ behaves like a
reproducing kernel, exactly as the ordinary Jordan--Pauli propagator,
which reproduces any solution of the KG equation, and produces one
such from an arbitrary spacetime function. 

Notice moreover that $\wt D$ is \textit{causal}: $D = 0$ when
$(x - x')^2 < 0$. This does not contradict Yngvason's
theorem~\cite{Yngvason70} on the nonlocality of quantum fields
associated to WPs, for, among other reasons, the wavefunctions depend
on an extra variable.

\section{Connecting with the standard formalism}
\label{sec:rear-view}

The \textit{point de d\'epart} of the standard formalism for the
Wigner modules is the choice of a reference 4-momentum 
$k = (|\kk|,\kk)$, which for massless particles can only be arbitrary.
Its ``length'' $|\kk|$ is irrelevant, so here it is assumed equal to
one. The time-honoured choice for the reference momentum is
$k := (1,0,0,1)$. The representation space of its corresponding little
group for a boson~WP is spanned by vectors lying on the circle
$|\xii|^2 := (\xi^1)^2 + (\xi^2)^2 = \ka^2$: either
$$
\ket{\xi^1,\xi^2} \equiv \ket{\ka;\tau},
\word{where} \tau := \arctan(\xi^2/\xi^1),
$$
or $\ket{\ka;h}$, with $h$ denoting the helicity, computed with
respect to the reference momentum. For these kets:
\begin{align*}
\bT_{1,2} \ket{\xi^1,\xi^2} 
&\equiv \bW^{1,2}\ket{\xi^1,\xi^2} = \xi^{1,2} \ket{\xi^1,\xi^2};
\\
\shortintertext{and also:}
\exp(i\bt\bW^0) \ket{\ka;\tau} &= \ket{\ka;\tau - \bt},
\word{or} \exp(i\bt\bW^0) \ket{\ka;h} = e^{i\bt h} \ket{\ka;h}.
\end{align*}
Then one can employ the \textit{standard} wave functions:
$$
\psi_\stan(\pp,\xi^1,\xi^2) := \braket{\pp,\xi^1,\xi^2}{\psi}
$$
defined on the lightcone and the internal circle by the customary
lifting to a unirrep space of the Poincar\'e group.

For a general unit vector $\kk$, the generators of rotations take the
form
\begin{equation}
\bL_{\kk} = -i\pp \x \del_\pp 
+ \frac{\pp \x (\kk \x \pp)}{|\pp|(|\pp| + \kk\.\pp)}\, \bS\.\kk
+ \frac{\pp}{|\pp|}\, \bS_\xii\.\kk
= -i \pp \x \del_\pp 
+ \frac{\pp + |\pp|\kk}{|\pp| + \kk\.\pp}\, \bS_\xii\.\kk
\label{eq:full-generator-L} 
\end{equation}
where $\xii$ is taken transversal to~$\kk$ of norm~$\ka$, and
$\bS_\xii := -i \xii \x \del_\xii$. For the boost generators,
one finds:
\begin{align}
\bK_{\kk} 
&= i|\pp|\,\del_\pp 
- \frac{\kk \x \pp}{|\pp| + \kk\.\pp}\, \bS_\xii\.\kk
+ \frac{\pp\.\xii}{|\pp|^2}\,\frac{\pp + |\pp|\kk}{|\pp| + \kk\.\pp}
- \frac{\xii}{|\pp|}
\notag \\
&= i|\pp|\,\del_\pp 
- \frac{\kk \x \pp}{|\pp| + \kk\.\pp}\, \bS_\xii\.\kk
+ \frac{\pp}{|\pp|^2} \x \biggl(
\frac{\pp + |\pp|\kk}{|\pp| + \kk\.\pp} \x \xii \biggr).
\label{eq:full-generator-K} 
\end{align}
The generators are defined on a (dense) subspace of the Hilbert space
consisting of twice-differentiable functions vanishing on a cylinder
centered on the negative $\kk$-axis, including the origin -- keep in
mind the analysis in~\cite{FlatoSF83}. When $\kk = (0,0,1)$, one
recovers from Eqs.~\eqref{eq:full-generator-L}
and~\eqref{eq:full-generator-K} the familiar expressions found by
Lomont and Moses~\cite{LomontM62} long ago.

It stands to reason that wavefunctions pertaining to the standard
routine must be related to the invariant wavefunctions of
Sect.~\ref{sec:hic-jacet-lepus} by unitary transformations. Let
$$
\al := \arccos(\kk\.\pp/|\pp|) 
= \arctan\frac{|\pp - (\pp\.\kk) \kk|}{\pp\.\kk}\,.
$$
In~\cite{Hirata77} one finds the assertion that such unitary
transformations essentially consist of a rotation representative:
$$
\dl(|\xii|^2 - \ka^2) \,\dl(\xii\.\kk) \,\psi_0(\pp,\xii) 
:= e^{iw^0/|\pp|} \exp\biggl(
i\al\,\frac{\kk \x \pp}{|\kk \x \pp|} \. \bS \biggr)
\Phi(\pp,w) \Bigr|_{\ww=\xii + w^0\pp/|\pp|} \,.
$$
Reciprocally, given $\kk$:
\begin{align*}
\Phi(\pp,w) = e^{-iw^0/|\pp|} \exp\biggl(
-i\al\,\frac{\kk \x \pp}{|\kk \x \pp|}\. \bS_\xii \biggr)
\,\dl(|\xii|^2 - \ka^2) \,\dl(\xii\.\kk)\, \psi_0(\pp,\xi)
\Bigr|_{\xii=\ww - w^0\pp/|\pp|}.
\end{align*}
Let us simply denote
$$
V := \exp\biggl( i\al\,\frac{\kk \x \pp}{|\kk \x \pp|} \. \bS \biggr).
$$
It is perfectly true that $V$ ``diagonalizes'' the helicity operator:
$$
V (\bS\.\pp/|\pp|) V^\7 = \bS\.\kk.
$$
Straightforward albeit tedious calculations show that the correct
internal angular momentum components transversal to $\kk$ in
\eqref{eq:full-generator-L} are recovered by this unitary 
transformation. (See also \cite{NiedererOR74, AsoreyBC85}.)
Unfortunately, we cannot go into this matter here.

\appendix

\section{Poincar\'e group conventions}
\label{app:back-to-basics}

Our metric on the Minkowski space $\bM$ is mostly-negative. The inner
product of two vectors $x \equiv x^\mu$, $p \equiv p^\nu$ of spacetime
is denoted with parentheses: $(xp) = x^\mu p_\mu$. When (we hope) it
does not cause confusion, we often write $p^2 = (pp)$, say.

The Lie algebra $\pl$ of $\sP$ has a basis of ten elements
$\{P^0,P^a,L^a,K^a : a = 1,2,3\}$, corresponding respectively to
time translations, space translations, rotations and boosts. The
commutation relations for the Lorentz subgroup are as follows:
\begin{align*}
[L^a,L^b] = \epsi^{ab}_c\, L^c,  \qquad 
[L^a,K^b] = \epsi^{ab}_c\, K^c,  \qquad
[K^a,K^b] = -\epsi^{ab}_c\, L^c.
\end{align*}
The commutation relations are realized%
\footnote{Or by $K^a = -\half \sg^a$ and $L^a = -\ihalf \sg^a$. In the
usual terminology, $K^a = \half\sg^a$ and $K^a = -\half\sg^a$
correspond to the $D(0,\half)$ and $D(\half,0)$ spinor
representations respectively, according to~\cite[Chap.~8]{SexlU01}.}
by $K^a = \half \sg^a$ and $L^a = -\ihalf \sg^a$.

In the real four-dimensional representation:
\begin{align*}
J^{01} &\equiv K^1 
= \begin{pmatrix} & 1 && \\ 1 &&& \\ && 0 & \\ &&& 0 \end{pmatrix}; 
\quad
J^{02} \equiv K^2 
= \begin{pmatrix} && 1 & \\ & 0 && \\ 1 &&& \\ &&& 0 \end{pmatrix}; 
\quad
J^{03} \equiv K^3 
= \begin{pmatrix} &&& 1 \\ && 0 & \\ & 0 && \\ 1 &&& \end{pmatrix};
\\
J^{23} &\equiv L^1 
= \begin{pmatrix} 0 &&& \\ & 0 && \\ &&& -1 \\ && 1 & \end{pmatrix};
\quad
J^{31} \equiv L^2 
= \begin{pmatrix} 0 &&& \\ &&& 1 \\ && 0 & \\ & -1 && \end{pmatrix};
\quad
J^{12} \equiv L^3 
= \begin{pmatrix} 0 &&& \\ && -1 & \\ & 1 && \\ &&& 0 \end{pmatrix},
\end{align*}
with the same commutation relations. Remark that
\begin{align*}
(L^1 + K^2)^2 
= \begin{pmatrix} 1 &&& -1 \\ & 0 && \\
&& 0 & \\ 1 &&& -1 \end{pmatrix} = (L^2 - K^1)^2
\end{align*}
and $(L^1 + K^2)^3 = (L^2 - K^1)^3 = 0$.

\goodbreak 

It is advisable to pull these generators together in matrix form:
$$
J^{\mu\nu} 
= \begin{pmatrix} & K^1 & K^2 & K^3 \\ -K^1 && L^3 & -L^2 \\
-K^2 & - L^3 && L^1 \\ -K^3 & L^2 & -L^1 & \end{pmatrix} \word{or}
J_{\mu\nu} 
= \begin{pmatrix} & -K^1 & -K^2 & -K^3 \\ K^1 && L^3 & -L^2 \\
K^2 & - L^3 && L^1 \\ K^3 & L^2 & -L^1 & \end{pmatrix}.
$$
The general expression is $(J_{\rho\sg})^\al_{\;\bt} 
= \dl^\al_\rho \,g_{\sg\bt} - \dl^\al_\sg \,g_{\rho\bt}$, and the
commutation relations are summarized as:
\begin{align}
[J_{\rho\sg}, J_{\mu\nu}] 
= -g_{\rho\mu}J_{\sg\nu} - g_{\sg\nu} J_{\rho\mu} 
+ g_{\sg\mu}J_{\rho\nu} + g_{\rho\nu}J_{\sg\mu}.
\label{eq:unmoved-movers} 
\end{align}

The dual tensor:
$$
J^{*\rho\mu} := -\half \eps^{\rho\mu\nu\tau} J_{\nu\tau}
= \begin{pmatrix} & -L^1 & -L^2 & -L^3 \\ L^1 && K^3 & -K^2 \\
L^2 & - K^3 && K^1 \\ L^3 & K^2 & -K^1 & \end{pmatrix}
$$
plays a role in the theory of the WP. Notice that 
$\KK\.\LL = \half J_{\rho\mu} J^{*\rho\mu}$ is a relativistic
invariant; as is $\KK^2 - \LL^2 = \half J_{\rho\mu} J^{\rho\mu} 
= -\half J^*_{\rho\mu} J^{*\rho\mu}$. These are just the Casimirs of
the Lorentz group. A generic infinitesimal Lorentz transformation is
of the form
$$
\La \simeq 1 + \half\om^{\rho\sg}J_{\rho\sg} \,, \word{or} 
\lLa^\mu_\nu = \dl^\mu_\nu + \om^\mu{}_\nu \,,
$$
where $\om^{\rho\sg}$ must be skewsymmetric. 

The $P^\mu$ mutually commute. The remaining nonvanishing commutation
relations for $\sP$ are given by:
$$
[L^a, P^b] = \epsi^{ab}_c\,P^c, \quad 
[K^a, P^b] = -\dl^{ab} P^0, \quad
[K^a, P^0] = -P^a;
$$
that is, 
$[J^{\ka\rho}, P^\mu] = g^{\mu\rho} P^\ka - g^{\mu\ka} P^\rho$.

Let $U(\La)$ be the unitary operator acting on one-particle states,
corresponding to a Lorentz transformation~$\La$. As discussed for
instance in \cite[Sect.~2.4]{Weinberg95I}, one finds that
$$
U^\7(\La) \,\bP^\mu\, U(\La) = \lLa^\mu_\nu \,\bP^\nu; \quad
U^\7(\La) \,\bJ^{\mu\nu}\, U(\La) 
= \lLa^\mu_\rho \lLa^\nu_\sg \,\bJ^{\rho\sg},
$$
where by $\bP$ and $\bJ = \{\bK,\bL\}$ we denote \textit{hermitian}
generators on Hilbert space, with commutation relations:
$$
[\bL^a, \bL^b] = i\epsi^{ab}_c\, \bL^c; \quad 
[\bL^a, \bK^b] = i\epsi^{ab}_c\, \bK^c; \quad
[\bK^a, \bK^b] = -i\epsi^{ab}_c\, \bL^c;
$$
that is, equation \eqref{eq:unmoved-movers} leads to
$$
[\bJ_{\rho\sg}, \bJ_{\mu\nu}] 
= i\bigl( -g_{\rho\mu} \bJ_{\sg\nu} - g_{\sg\nu} \bJ_{\rho\mu}
+ g_{\sg\mu} \bJ_{\rho\nu} + g_{\rho\nu} \bJ_{\sg\mu} \bigr).
$$

\section{The Lorentz decompositions of null rotations}
\label{app:nullius-in-verba}

The unique decomposition of an arbitrary (proper orthochronous)
Lorentz matrix~$S$ into the product of a rotation and a boost is well
known~\cite[Ch.~1]{SexlU01}. It becomes
$$
S = \twobytwo{\al}{\aaa^t}{\cc}{N}
= \twobytwo{1}{0}{0}{N - \cc\aaa^t/(1 + \al)} L_{\aaa/\al}
=: \twobytwo{1}{0}{0}{N - \cc\aaa^t/(1 + \al)}
\twobytwo{\al}{\aaa^t}{\aaa}{1_3 + \frac{\aaa\aaa^t}{1 + \al}},
$$
where $\al^2 = 1 + \aaa^2$. Since $S$ and $S^t$ are Lorentz, which
implies $N\aaa = \al\cc$, $N^t\cc = \al\aaa$ and
$N^t N = 1_3 + \aaa\aaa^t$, one checks that 
$R := N - \cc\aaa^t/(1 + \al)$ is a rotation and that $R\aaa = \cc$,
and thus also $R + R\aaa\aaa^t/(1 + \al) = N$.

We want to decompose null rotations in~$G_p$. Note that there is an
infinity of spacelike surfaces, of timelike, null or spacelike
vectors, which are orbits of~$G_p$ in~$\bM$, each isometric to the
group of motions of a plane~\cite{Mund07}. Consider those null
rotations which leave invariant the standard momentum $k = (1,0,0,1)$.
Denoting when convenient $b_1^2 + b_2^2$ by $|b|^2$, a general null
rotation fixing $k$ is given by:
$$
S(b_1,b_2) := \begin{pmatrix} 
1 + \half|b|^2 & -b_2 & b_1 & -\half|b|^2 \\
-b_2 & 1 & 0 & b_2 \\
b_1 & 0 & 1 & -b_1 \\
\half|b|^2 & -b_2 & b_1 & 1 - \half|b|^2 \end{pmatrix}
=: \twobytwo{\al}{\aaa^t}{\cc}{N}.
$$

Simplifying further, we work out first the case $S(0,-b)$, with 
$b > 0$. 

Here $\al^2 = 1 + b^2 + \quarter b^4 = (1 + \half b^2)^2$ so that 
$1 + \al = \half(4 + b^2)$, and $S(0,-b)$ factorizes as
\begin{align*}
& \begin{pmatrix} 
1 + \half b^2 & b & 0 & -\half b^2 \\
b & 1 & 0 & -b \\
0 & 0 & 1 & 0 \\
\half b^2 & b & 0 & 1 - \half b^2 \end{pmatrix}
= \begin{pmatrix}
1 & 0 & 0 & 0 \\
0 & \frac{4 - b^2}{4 + b^2} & 0 & -\frac{4b}{4 + b^2} \\
0 & 0 & 1 & 0 \\
0 & \frac{4b}{4 + b^2} & 0 & \frac{4 - b^2}{4 + b^2}
\end{pmatrix}
\begin{pmatrix}
1 + \half b^2 & b & 0 & -\half b^2 \\
b & 1 + \frac{2b^2}{4+b^2} & 0 & -\frac{b^3}{4+b^2} \\
0 & 0 & 1 & 0 \\
-\half b^2 & -\frac{b^3}{4+b^2} & 0 & 1 + \frac{b^4}{2(4+b^2)}
\end{pmatrix}
\\[\jot]
&\quad =: RL = (R L R^{-1}) R =: L'R.
\end{align*}
We see clearly that $R$ is a rotation around the $y$-axis, of positive
angle $\th$ turning anticlockwise from the positive $z$-axis towards
the positive $x$-axis, with $\th = 2\arctan(b/2)$. The velocity
associated with the boost $L'$ is:
$$
\vv = \bigl( 2b/(2+b^2), 0, -b^2/(2+b^2) \bigr);
$$
therefore its rapidity parameter is given by 
$\ze = \arcsinh \bigl( \half b\sqrt{4 + b^2} \bigr)$; the direction of
the boost forms an angle $\arctan(b/2)$ with the $x$-axis, tilted
towards the negative $z$-axis. For small angles, it is intuitive that
the boost undoes the turn effected by the rotation. The result
reproduces the one indicated without proof in~\cite{Yngvason70}.

\subsection*{Acknowledgements}

A report by Alejandro Jenkins of a conversation with Mark Wise set
this work in motion. We are grateful to Alejandro, as well as to
Fedele Lizzi and Patrizia Vitale, for discussions on the gyroscope
property, and to Karl-Henning Rehren for most useful remarks about the
equations by Wigner. We thank Daniel Sol\'is for checking
App.~\ref{app:nullius-in-verba} and a useful observation.

The project has received funding from the European Union's
Horizon~2020 research and innovation programme under the Marie
Sk{\l}odowska-Curie grant agreement No.~690575. 
JMG-B received funding from Project FPA2015--65745--P of MINECO/Feder,
and acknowledges the support of the COST action QSPACE.
JCV received support from the Vicerrector\'ia de Investigaci\'on of
the Universidad de Costa~Rica.

\newpage 


\begin{thebibliography}{37}

\footnotesize

\bibitem{Wigner39}
E. P. Wigner,
``On unitary representations of the inhomogeneous Lorentz group'',
Ann. Math. \textbf{40} (1939), 149--204.

\bibitem{Schroer17}
B. Schroer,
``Wigner's infinite spin representations and inert matter'',
Eur. Phys. J. C\,77:362 (2017).

\bibitem{SToro13a}
Ph. Schuster and N. Toro,
``On the theory of continuous spin particles: wavefunctions and
soft-factor scattering amplitudes'',
JHEP \textbf{1309} (2013), 104.

\bibitem{SToro13b}
Ph. Schuster and N. Toro,
``On the theory of continuous-spin particles: helicity correspondence
in radiation and forces'',
JHEP \textbf{1309} (2013), 105.

\bibitem{SToro15}
Ph. Schuster and N. Toro,
``Continuous-spin particle field theory with helicity 
correspondence'',
Phys. Rev. D \textbf{91} (2015), 025023.

\bibitem{BekaertNS16}
X. Bekaert, M. Najafizadeh and M. R. Setare,
``A gauge field theory of fermionic continuous-spin particles'',
Phys. Lett. B \textbf{760} (2016), 320--323.

\bibitem{BekaertS17}
X. Bekaert and E. Skvortsov,
``Elementary particles with continuous spin'',
Int. J. Mod. Phys. A \textbf{32} (2017), 1730019.

\bibitem{Najafizadeh18}
M. Najafizadeh,
``Modified Wigner equations and continuous spin gauge field'',
Phys. Rev. D \textbf{97} (2018), 065009.

\bibitem{KhabarovZ18}
M. V. Khabarov and Yu. M. Zinoviev,
``Infinite (continuous) spin fields in the frame-like formalism'',
Nucl. Phys. B \textbf{928} (2018), 182--216.

\bibitem{BuchbinderKT18}
I. L. Buchbinder, V. A. Krykhtin and H.Takata,
``BRST approach to Lagrangian construction for bosonic continuous spin
field'',
Phys. Lett. B \textbf{785} (2018), 315--319.

\bibitem{Rehren17}
K.-H. Rehren,
``Pauli--Luba\'nski limit and stress-energy tensor for infinite-spin
fields'',
JHEP \textbf{1711} (2017), 130.

\bibitem{MundRS17}
J. Mund, K.-H. Rehren and B. Schroer,
``Helicity decoupling in the massless limit of massive tensor fields'',
Nucl. Phys. B \textbf{924} (2017), 699--727.

\bibitem{Euterpe}
J. M. Gracia--Bond\'ia, F. Lizzi, J. C. V\'arilly and P. Vitale,
``The Kirillov picture for the Wigner particle'',
J. Phys. A. \textbf{51} (2018), 255203.

\bibitem{McKerrell65}
A. McKerrell,
``Canonical representations for massless particles and zero-mass 
limits of the helicity representation'',
Proc. Roy. Soc. London A \textbf{285} (1965), 287--296.

\bibitem{LomontM67}
J. S. Lomont and H. E. Moses,
``Reduction of reducible representations of the infinitesimal 
generators of the proper orthochronous inhomogeneous Lorentz group'',
J. Math. Phys. \textbf{8} (1967), 837--850.

\bibitem{Schwinger70}
J. Schwinger,
\textit{Particles, Sources and Fields}, vol.~1,
Addison-Wesley, Reading, MA, 1970.

\bibitem{Ganymede}
J. F. Cari\~nena, J. M. Gracia--Bond\'ia and J. C. V\'arilly,
``Relativistic quantum kinematics in the Moyal representation'',
J. Phys. A \textbf{23} (1990), 901--933.

\bibitem{BalachandranMSSZ92}
A. P. Balachandran, G. Marmo, A. Simoni, A. Stern and F. Zaccaria,
``On a classical description of massless particles'',
in \textit{Proceedings of the ISAQTP--Shanxi},
ed. by J. Q. Liang, M. L. Wang, S. N. Qiao and D. C. Su (1993),
pp.~396--402.

\bibitem{SexlU01}
R. U. Sexl and H. K. Urbantke,
\textit{Relativity, Groups, Particles},
Springer, Vienna, 2001.

\bibitem{Wigner57}
E. P. Wigner,
``Relativistic invariance and quantum phenomena'',
Rev. Mod. Phys. \textbf{29} (1957), 255--268.

\bibitem{BoyaCS74}
 L. J. Boya, J. F. Cari\~nena and M. Santander,
``On the continuity of the boosts for each orbit'',
Commun. Math. Phys. \textbf{37} (1974), 331--334.

\bibitem{FlatoSF83}
M. Flato, D. Sternheimer and C. Fr{\o}nsdal,
``Difficulties with massless particles?'',
Commun. Math. Phys. \textbf{90} (1983), 563--573.

\bibitem{BargmannW48}
V. Bargmann and E. P. Wigner, 
``Group theoretical discussion of relativistic wave equations'',
PNAS \textbf{34} (1948), 211--223.

\bibitem{Wigner48}
E. P. Wigner,
``Relativistische Wellengleichungen'',
Z. Physik \textbf{124} (1948), 665--684.

\bibitem{Wigner63}
E. P. Wigner,
``Invariant quantum mechanical equations of motion'',
in \textit{Theoretical Physics Lectures}, ed. by A. Salam,
International Atomic Energy Agency, Vienna, 1963; pp.~59--82.

\bibitem{BekaertM06}
X. Bekaert and J. Mourad,
``The continuous spin limit of higher spin field equations'',
JHEP \textbf{01} (2006) 115.

\bibitem{KHRprivate}
K.-H. Rehren, private communication.

\bibitem{SToro13c}
Ph. Schuster and N. Toro,
``A gauge field theory of continuous spin particles'',
JHEP \textbf{1310} (2013), 061.

\bibitem{Rivelles17}
V. O. Rivelles,
``Remarks on a gauge theory for continuous spin particles'',
Eur. Phys. J. C \textbf{77}:433 (2017).

\bibitem{PenroseR84}
R. Penrose and W. Rindler,
\textit{Spinors and Spacetime}, vol.~1,
Cambridge University Press, Cambridge, 1984.

\bibitem{Hirata77}
K. Hirata,
``Quantization of massless fields with continuous spin'',
Prog. Theor. Phys. \textbf{58} (1977) 652--666.

\bibitem{Yngvason70}
J. Yngvason,
``Zero-mass infinite spin representations of the Poincar\'e group
and quantum field theory'',
Commun. Math. Phys. \textbf{18} (1970), 195--203.

\bibitem{LomontM62}
J. S. Lomont and H. E. Moses,
``Simple realizations of the infinitesimal generators of the proper 
orthochronous inhomogeneous Lorentz group for mass zero'',
J. Math. Phys. \textbf{3} (1962), 405--408.

\bibitem{NiedererOR74}
U. H. Niederer and L. O'Raifeartaigh,
``Realizations of the unitary representations of the inhomogeneous
space-time groups I'',
Fortschr. Phys. \textbf{22} (1974), 111--129.

\bibitem{AsoreyBC85}
M. Asorey, L. J. Boya and J. F. Cari\~nena,
``Covariant representations in a fibre bundle framework'', 
Rep. Math. Phys. \textbf{21} (1985), 391--404. 

\bibitem{Weinberg95I}
S. Weinberg,
\textit{The Quantum Theory of Fields I},
Cambridge University Press, Cambridge, 1995.

\bibitem{Mund07}
J. Mund,
``String-localized covariant quantum fields'',
in \textit{Rigorous Quantum Field Theory}, ed. by
A. Boutet de~Monvel, D. Buchholz, D. Iagolnitzer and U.~Moschella
(Birkh\"auser, Boston, 2007), pp.~199--212.

\end{thebibliography}
\end{document}